\documentclass[a4paper,12pt]{article}
\usepackage{textgreek}
\usepackage[T1]{fontenc}
\usepackage[utf8]{inputenc}
\usepackage{graphicx}
\usepackage{amssymb,amsmath,amsthm}
\usepackage{subfigure}
\usepackage[margin=2.5cm]{geometry}
\usepackage[detect-all]{siunitx}
\usepackage[version=4]{mhchem}
\usepackage{authblk}
\usepackage{tikz}
\usetikzlibrary{calc,patterns,angles,quotes} 
\usepackage[hidelinks]{hyperref}

\DeclareSIUnit{\Molar}{M}

\graphicspath{{figs/}}

\title{Liquid Marble Photosensor}

\author[1]{Andrew Adamatzky}
\author[1]{Michail-Antisthenis Tsompanas}
\author[1]{\\ Thomas C.\ Draper}
\author[1,2]{Richard Mayne}
\affil[1]{Unconventional Computing Laboratory, University of the West of England,
Coldharbour Lane, Bristol, BS16 1QY, UK}
\affil[2]{Department of Applied Sciences,
University of the West of England,
Coldharbour Lane, Bristol, BS16 1QY, UK}
\date{}
\begin{document}

\maketitle

\begin{abstract}
\noindent
A liquid marble is a liquid droplet coated by a hydrophobic power. The liquid marble does not wet adjacent surfaces and therefore can be manipulated as a dry soft body. A Belousov-Zhabotinsky (BZ) reaction is an oscillatory chemical reaction exhibiting waves of oxidation. We demonstrate how to make a photo-sensor from BZ medium liquid marbles. We insert electrodes into a liquid marble, prepared from BZ solution and coated with polyethylene powder. The electrodes record a potential difference which oscillates due to oxidation wave-fronts crossing the electrodes. When the BZ marble is illuminated by a light source, the oxidation wave-fronts are hindered and, thus, the electrical potential recorded ceases to oscillate. We characterise several types of responses of BZ marble photosensors to various stimuli, and provide explanations of the recorded activity. BZ liquid marble photosensors may find applications in the fields of liquid electronics, soft robotics and unconventional computing. 

\vspace{3mm}
\end{abstract}

\section{Introduction}

A non-stirred Belousov-Zhabotinsky (BZ) medium~\cite{belousov1959periodic, zhabotinsky1964periodic} exhibits a rich spectrum of oxidation wave-front dynamics. In late 1980s Kuhnert, Agladze and Krinsky~\cite{kuhnert1986new, kuhnert1989image} experimentally demonstrated implementation of image processing and memory in BZ media, where data was input optically and the results of the computation were represented by the patterns of the oxidation wave-fronts. Since then, a large variety of unconventional computing devices have been implemented using this BZ medium. These include diodes~\cite{DBLP:journals/ijuc/IgarashiG11}, logical gates~\cite{steinbock1996chemical}, robot controllers~\cite{adamatzky2004experimental,tsompanas2019belousov},  counters~\cite{gorecki2003chemical}, neuromorphic architectures~\cite{gorecki2006information,gentili2012belousov,gruenert2015understanding, stovold2017associative} and arithmetical circuits~\cite{suncrossover,stevens2012time}. Methods for optical, geometrical or chemical control of oscillation frequency have been employed to prototype logical gates~\cite{gorecki2014information}, modulators~\cite{sielewiesiuk2002passive}, filters~\cite{gorecka2003t}, memory~\cite{gizynski2017chemical}, fuzzy logic~\cite{gentili2012belousov} and oscillatory associated memory~\cite{calayir2013fully}. Frequency of oscillations in the BZ reaction is the driving force behind the computation in BZ medium, thus being able to control the frequency of oscillations is important and is one of the key motivations for the present paper. 

In order to prototype a fully fledged computer via BZ medium, we need to encapsulate and isolate parts of the medium, e.g.\ following a theoretical ideology of membrane computing, or P-systems, introduced by P\u{a}un in 1998 \cite{DBLP:journals/eatcs/Paun99a,DBLP:journals/ijfcs/Paun00, Paun2002, Paun09}. Encapsulation can be achieved by immersing BZ droplets in oil and allowing pores to form between the droplets, thus permitting the exchange of information in a form of travelling oxidation wave-fronts. This has been demonstrated in \cite{vanag2001pattern,vanag2004waves,kaminaga2006reaction,szymanski2013chemo,gorecki2015chemical,wang2016configurable,gruenert2013multi,henson2015towards}. This approach however has its limitations, because the liquid `mother' solution (in which the BZ droplets are dispersed) requires special handling. An ideal approach would be to make BZ droplets `dry', yet being able to produce controllable oscillation dynamics. As a result, the required alternative solution comes in the form of liquid marbles. 

Liquid marbles (LMs)~\cite{aussillous2001liquid} are liquid droplets coated by hydrophobic particles at the liquid/air interface. The key characteristic of LMs is that they do not wet adjacent surfaces and can therefore be manipulated by a variety of means~\cite{ooi2015manipulation}: rolling, mechanical lifting and dropping, sliding and floating~\cite{bormashenko2009mechanism,draper2018a,celestini2018propulsion,draper2018ucnc}. The range of potential applications for LMs is huge and spans most fields of life sciences, chemistry, physics and engineering~\cite{bormashenko2011liquid,mchale2011liquid,draper2017,avruamescu2018liquid,daeneke2018liquid}. Previously,  we demonstrated BZ marbles support non-trivial patterns of chemical oscillations~\cite{fullarton2018belousov}. 

A key limitation for using BZ LMs as computing elements is the lack of definite optical indication of the oxidation wave fronts, especially if the coating is non-transparent. To overcome this limitation, we proposed to measure oscillations of electrical potential, caused by oxidation wave-fronts by inserting electrodes in the BZ medium-filled LMs. We previously made a prototype BZ marble thermal sensor~\cite{adamatzky2019thermal} and demonstrated that it is possible to reduce the frequency of oscillations or even completely halt oscillations by `freezing' them. This method is robust, but requires Peltier elements and, therefore, lacks the scalability demanded from a computational viewpoint. An alternative means for controlling oscillations in BZ marbles is illumination with visible light, as the BZ medium is proven to be a light-sensitive chemical system due to the photochemical properties of the catalyst~\cite{Gaspar1983,kuhnert1989image,Hanazaki1995,rambidi1998information,toth2000wave,toth2000wave,wang2010intelligent}. 

Based on this rationale, we decided to explore whether it is possible to prototype a BZ LM photosensor, which would react to illumination by the consequent changes in its oscillatory patterns of electrical potential. Our laboratory experiments and computer modelling conclusively demonstrated that BZ LMs are fully functional photosensors capable of responding to a series of optical stimulation. The paper is structured as follows. In Sect.~\ref{methods} we outline chemical protocols, experimental setup and numerical techniques used. We analyse responses of the BZ marbles to optical stimulation in Sect.~\ref{results}, followed by a brief summary in Sect.~\ref{summary}.

\section{Methods}
\label{methods}


The LMs in this report were all formed using the Belousov-Zhabotinsky (BZ) reaction medium as the aqueous core and ultra-high density polyethylene (PE) (Sigma-Aldrich, \numrange[range-phrase = --]{3}{6e6}~\si{\gram\per\mole}, grain size approximately \SI{100}{\micro\metre}) as the powder coating. LMs were produced by hand-rolling droplets of the pre-prepared BZ medium (\SI{62}{\micro\L}) on a powder bed of PE, until a uniformly coated LM was generated.

The BZ reaction medium was prepared using a modified version of a lit\-er\-a\-ture method, omitting the surfactant Triton-X~\cite{Field1979}. Sodium bromide, malonic acid, sodium bromate, and ferroin were sourced from Sigma-Aldrich. Sulphuric acid was sourced from Fischer Scientific. All reagents were used as received, without any further purification. Sodium bromate (\SI{5.0}{\g}) was added to sulphuric acid (\SI{0.5}{\Molar}, \SI{69}{\ml}) with stirring. An aliquot of this solution was taken (\SI{3.0}{\milli\L}), and the remainder stored for future use. To the aliquot was added malonic acid (\SI{1.0}{\Molar}, \SI{0.5}{\milli\L}) and sodium bromide (\SI{1.0}{\Molar}, \SI{0.25}{\milli\L}), resulting in a transiently orange solution caused by the emission of bromine gas. Once the reaction mixture became colourless once more, ferroin was added (\SI{0.025}{\Molar}, \SI{0.5}{\milli\L}, Sigma-Aldrich product 318922, see ref~\cite{Field1979}), yielding approximately \SI{4.25}{\milli\L} of the BZ reaction medium.

The produced BZ LMs were set on a Petri dish, where they were fixed in place by being punctured with two iridium-coated stainless steel sub-dermal needles with twisted cables (Spes Medica SRL, Italy), for use as electrodes. 
The electrical output of the LM was logged with a PicoLog ADC-24 high resolution data logger (Pico Technology Ltd, UK). 

\begin{figure}[!tbp]
    \centering
   \subfigure[]{\includegraphics[width=0.25\textwidth]{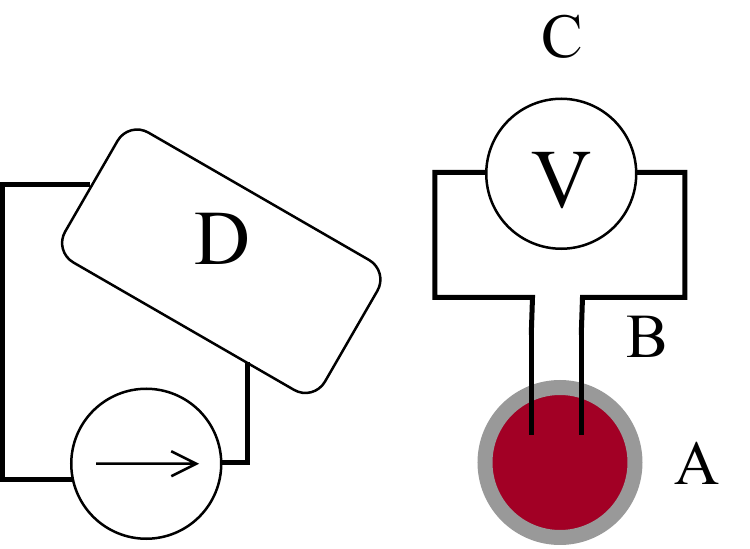}}
    \subfigure[]{\includegraphics[width=0.2\textwidth]{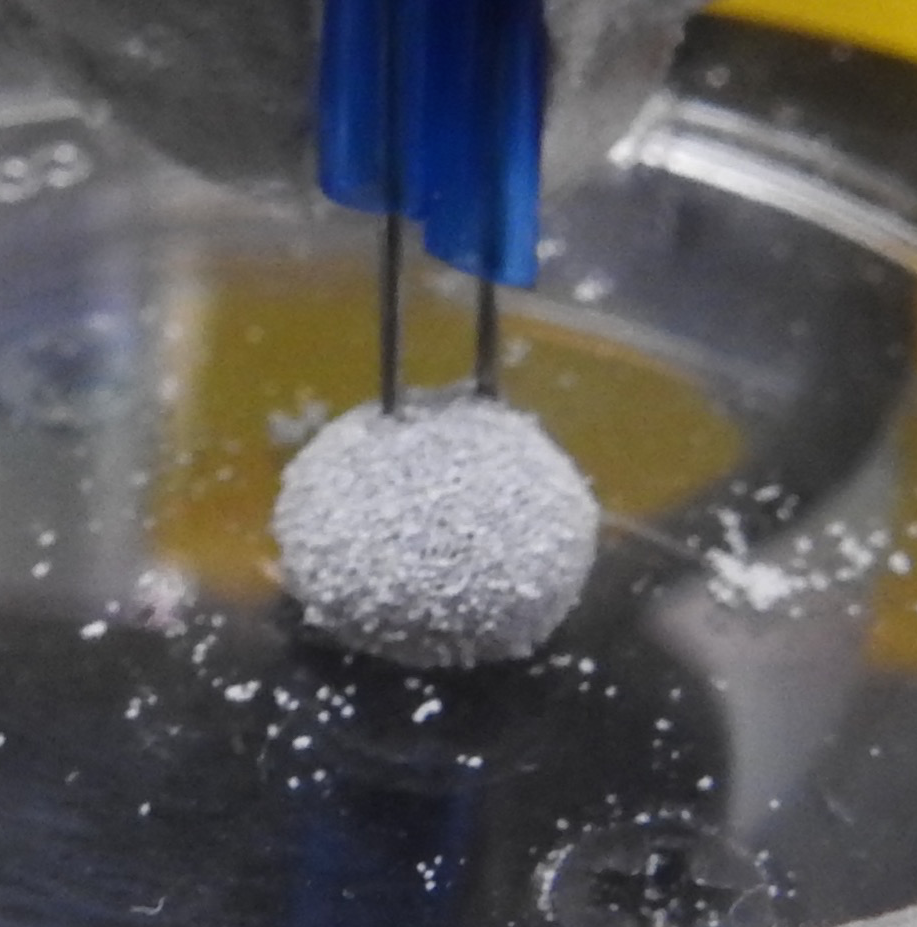}}
    \subfigure[]{\includegraphics[width=0.45\textwidth]{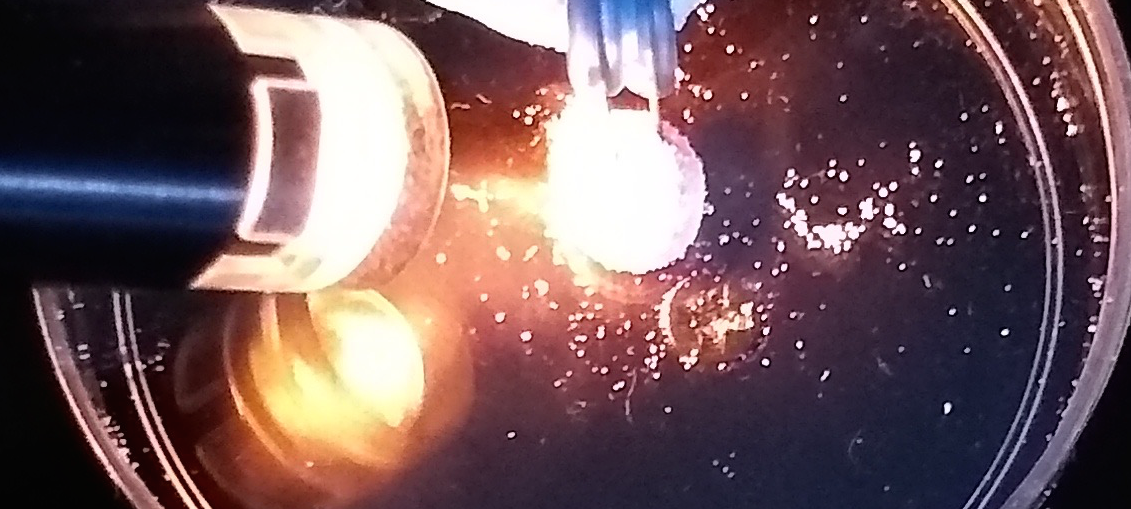}}
    \caption{Experimental setup. (a)~A scheme of the experimental setup: A -- BZ marble, B -- a pair of electrodes, C --- Pico ADC-24 logger, D -- light source.   (b)~A photo of BZ marble with electrodes insert. (c)~A photo of BZ marble illuminated (intensity of illumination is lower than in experiments, to prevent photo overexposure).}
    \label{fig:experimentalSetup}
\end{figure}

The BZ LMs were stimulated with a cold light source (PL2000, Photonic Optics, USA) 3250~K, 18~MLux, for 300~sec in average over experiments. A scheme of the experimental setup is shown in Fig.~\ref{fig:experimentalSetup}.

To explain patterns of electrical activity recorded, the propagation of waves in the LM was simulated by employing a numerical integration of a two-dimensional model of two-variable Oregonator equations~\cite{field1974oscillations,beato2003pulse}:

\begin{eqnarray}
  \frac{\partial u}{\partial t} & = & \frac{1}{\epsilon} (u - u^2 - (f v + \phi)\frac{u-q}{u+q}) + D_u \nabla^2 u \nonumber \\
  \frac{\partial v}{\partial t} & = & u - v.
\label{equ:oregonator}
\end{eqnarray}

 The variables $u$ and $v$ represent local concentrations of an activator, or an excitatory component of BZ system, and an inhibitor, or a refractory component. Parameter $\epsilon$ sets up a ratio of the time scale of variables $u$ and $v$; $q$ is a scaling parameter depending on rates of activation/propagation and inhibition; and $f$ is a stoichiometric coefficient. The system was integrated using the Euler method with five-node Laplace operator, time step $\Delta t=0.001$ and grid point spacing $\Delta x = 0.25$, $\epsilon=0.02$, $f=1.4$, $q=0.002$. The value of $\phi$ was varied from the interval $\Phi=[0.05,0.08]$, where constant $\phi$ is a rate of inhibitor production. The BZ LM was represented as a disc with a radius of 185 nodes, and the electrodes were represented as rectangular domains $\mathcal{E}_1$ and $\mathcal{E}_2$. The potential difference was calculated at each iteration $t$ as 
 $\sum_{x \in \mathcal{E}_2} u^t_x - \sum_{x \in \mathcal{E}_1} u^t_x$.

\section{Results}
\label{results}

\begin{figure}[!tbp]
    \centering
    \subfigure[]{\includegraphics[width=0.49\textwidth]{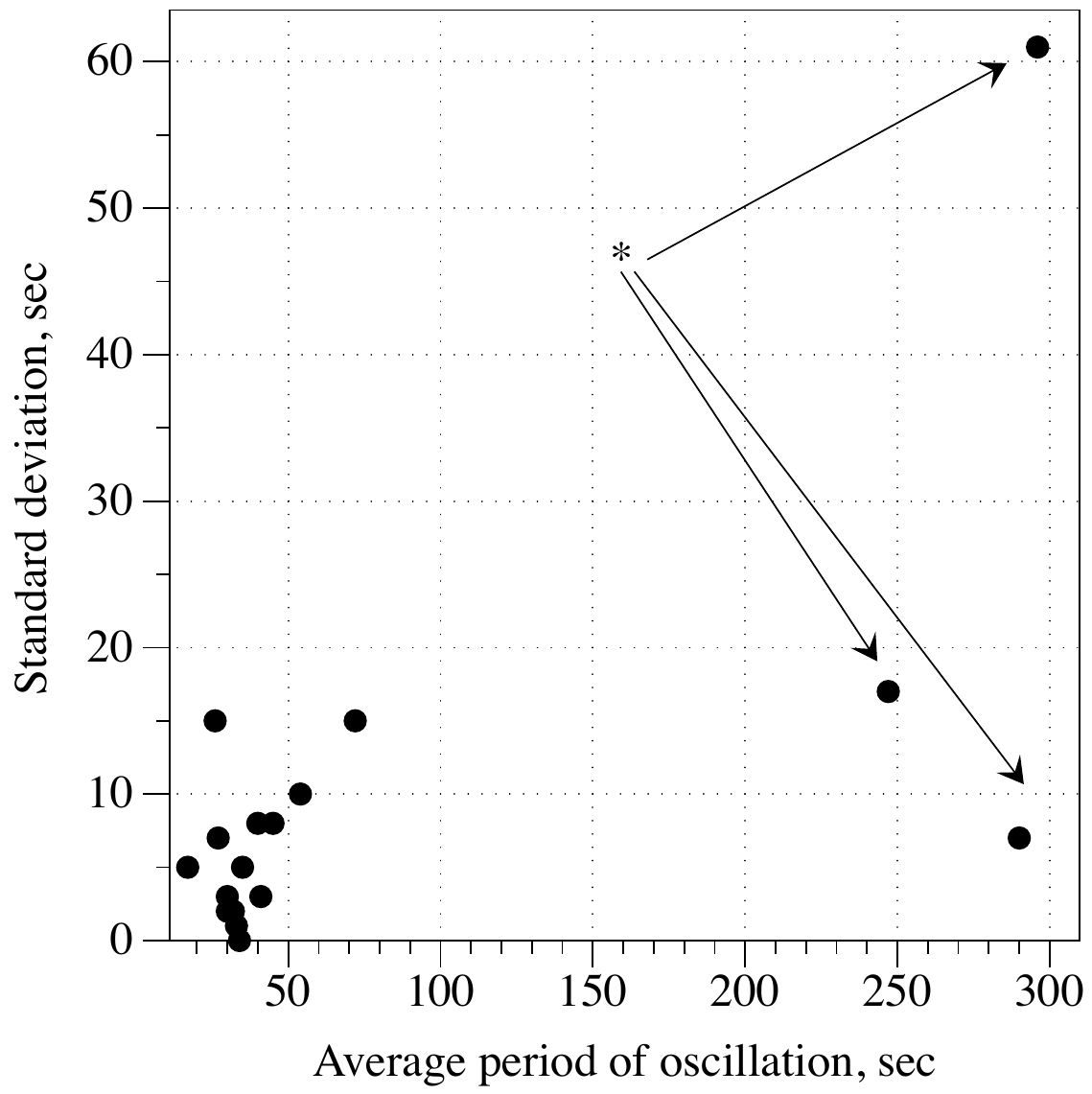}\label{fig:allperiods}}
    \subfigure[]{\includegraphics[width=0.49\textwidth]{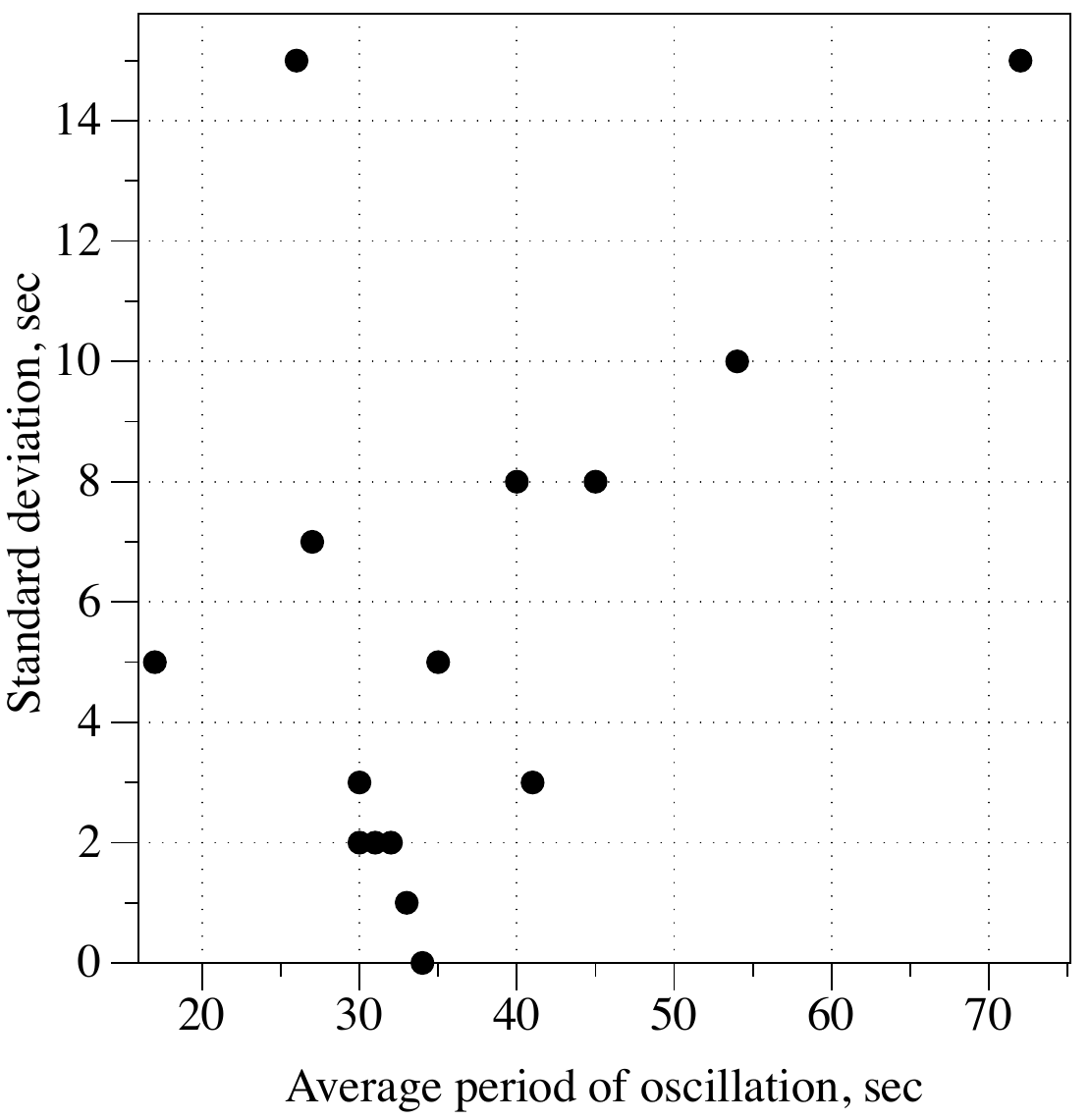}\label{fig:typicalperiods}}\\
    \subfigure[]{\includegraphics[width=0.49\textwidth]{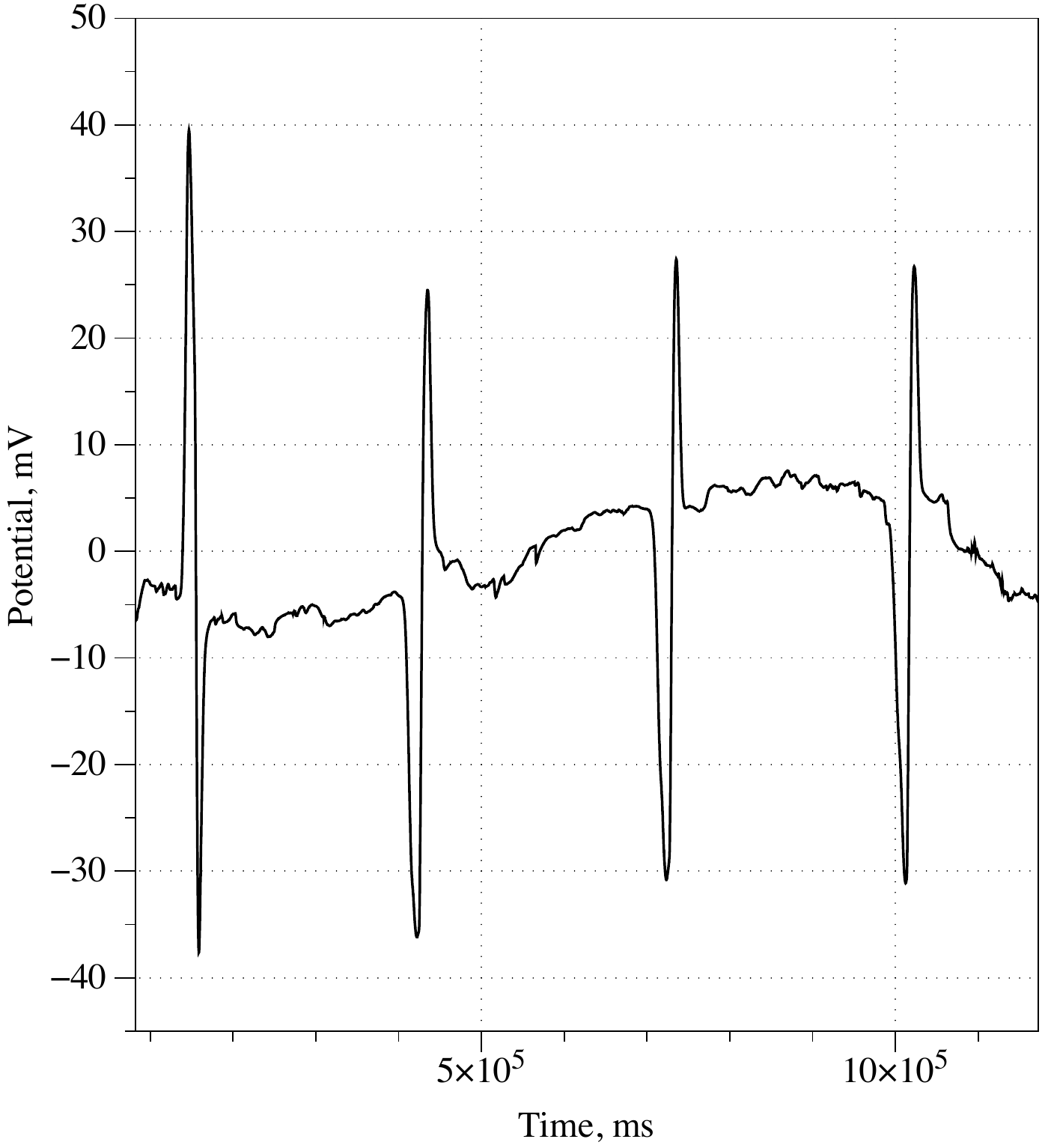}\label{fig:exampleLowFrequency}}
    \subfigure[]{\includegraphics[width=0.49\textwidth]{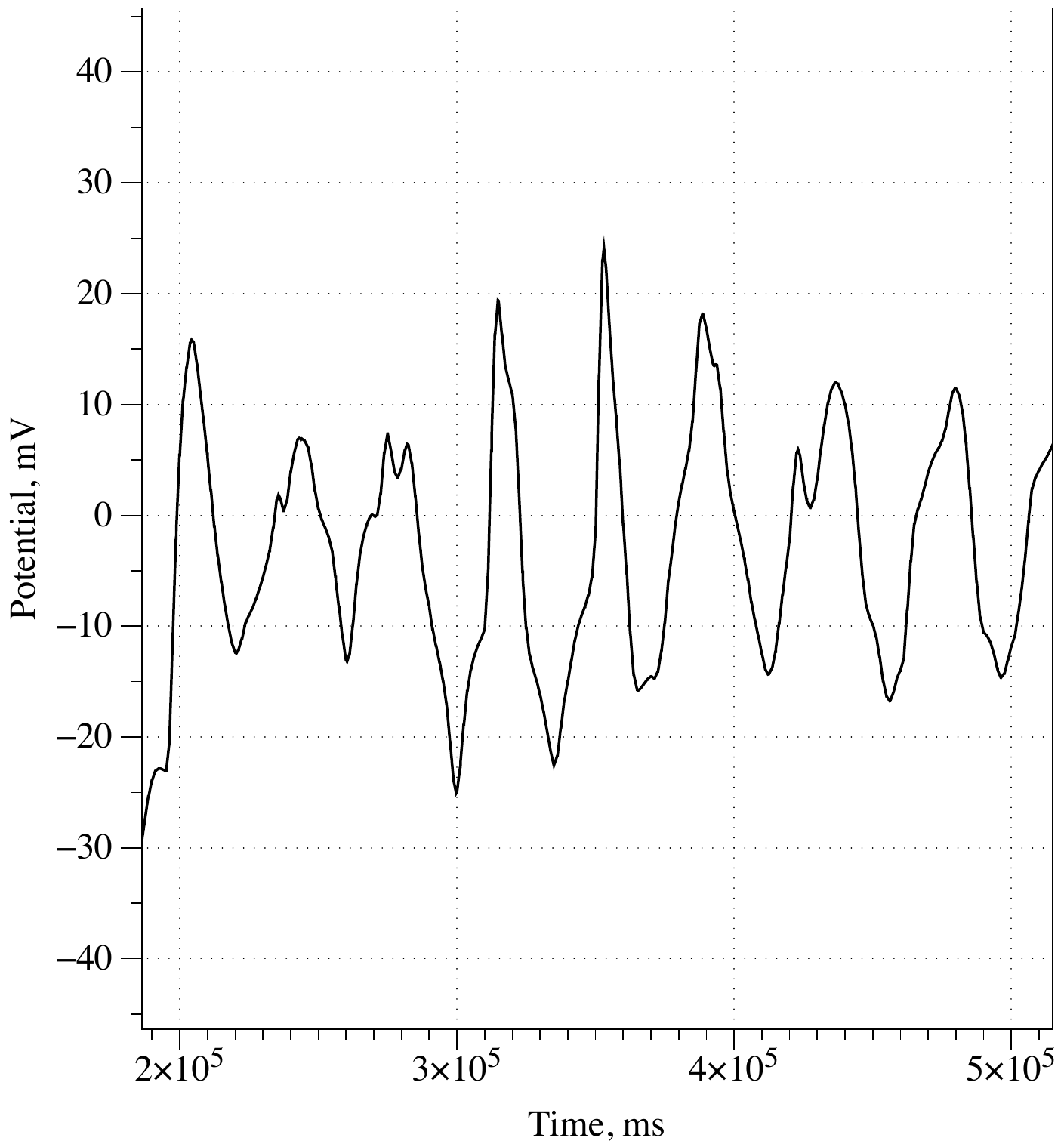}\label{fig:exampleTypicalFrequency}}
    \caption{Typology of oscillations. (a)~Average period of oscillations vs standard deviation for all marbles. Data points labelled with an asterisk are considered to be low-frequency liquid marbles. (b)~Zoomed in domain where majority of marbles are observed. Each dot on the plot is derived from experiments with unique marble. (c,d)~Exemplar experimental oscillations of (c)~low frequency marbles (shown in (a) by arrows) and (d)~high frequency marbles. }
    \label{fig:averageperiodsIntact}
\end{figure}

Average periods of oscillations for 19 marbles, recorded before stimulation with light, together with standard deviations ($\sigma$) for the respective LMs, are shown in Fig.~\ref{fig:allperiods}. Each dot corresponds to a particular LM. Most dots are concentrated in the domain \mbox{[0,70]\,s $\times$ [0,15]\,s}. 
The three dots indicated by arrows and an asterisk in Fig.~\ref{fig:allperiods} correspond to periods of oscillations over 240~s. The periods are  247~s ($\sigma$=17), 290~s ($\sigma$=7), and 298~s ($\sigma$=61). An example of low frequency oscillations recorded from a BZ LM is shown in Fig.~\ref{fig:exampleLowFrequency}.

\subsection{Low frequency liquid marbles}

\begin{figure}[!tbp]
    \centering
   \subfigure[$t=60$]{\includegraphics[width=0.21\textwidth]{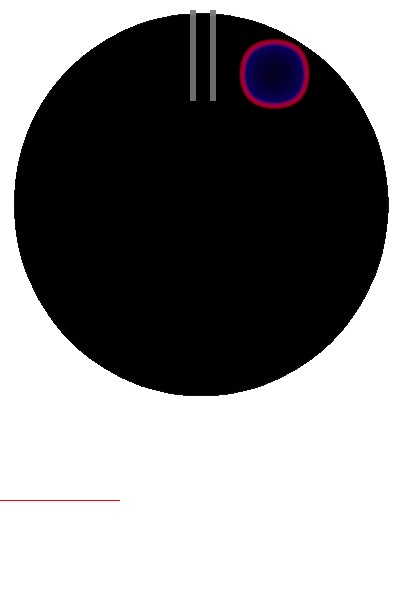}} 
    \subfigure[$t=1150$]{\includegraphics[width=0.21\textwidth]{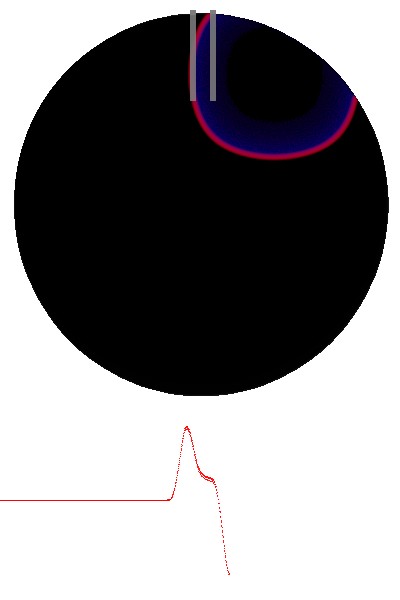}}
    \subfigure[$t=1300$]{\includegraphics[width=0.21\textwidth]{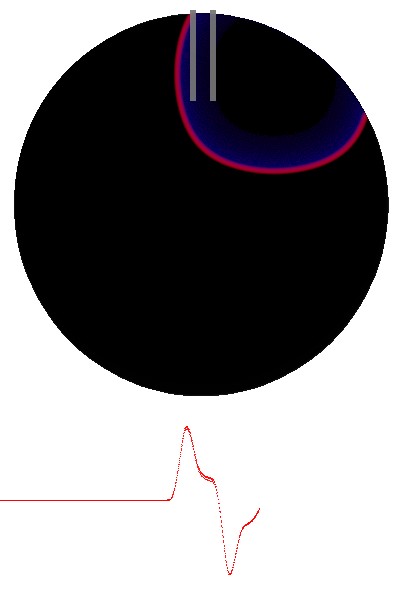}}
    \subfigure[$t=2000$]{\includegraphics[width=0.21\textwidth]{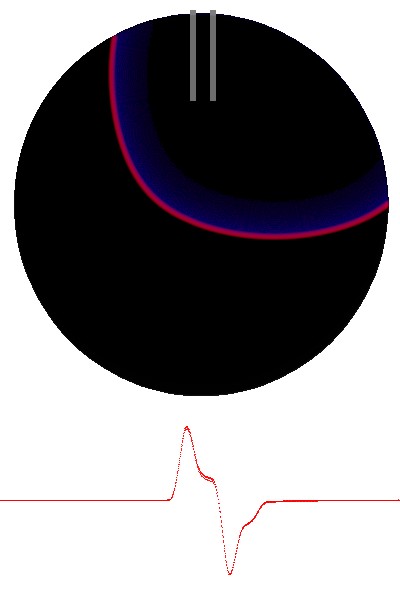}}
    \subfigure[$t=3100$]{\includegraphics[width=0.21\textwidth]{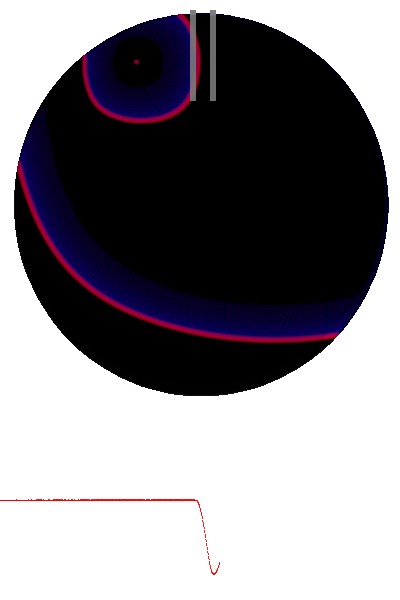}}
    \subfigure[$t=3450$]{\includegraphics[width=0.21\textwidth]{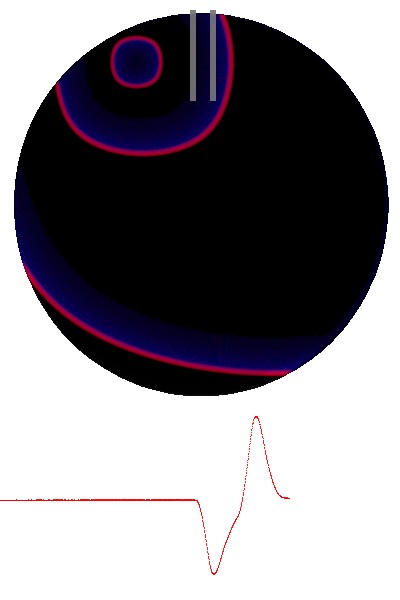}}
    \subfigure[$t=3850$]{\includegraphics[width=0.21\textwidth]{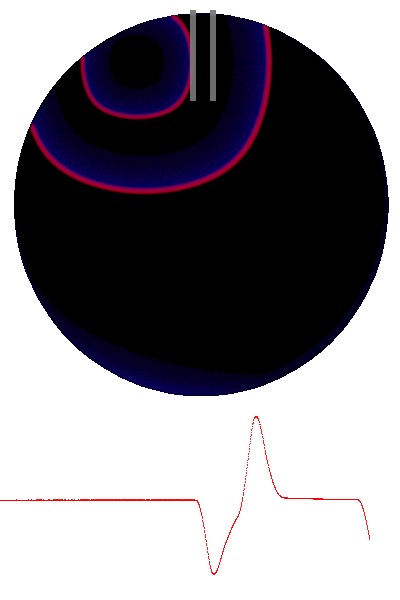}}
    \subfigure[$t=5950$]{\includegraphics[width=0.21\textwidth]{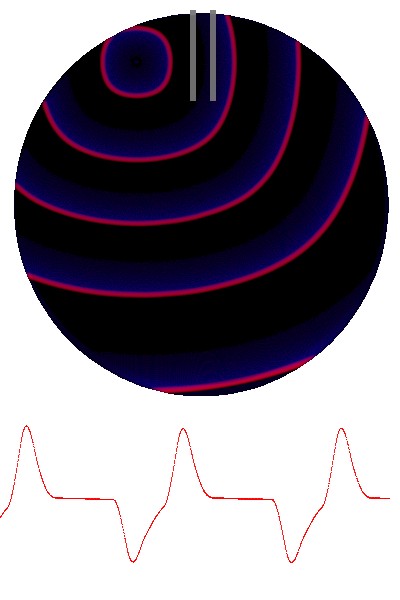}}
    \subfigure[]{\includegraphics[width=0.56\textwidth]{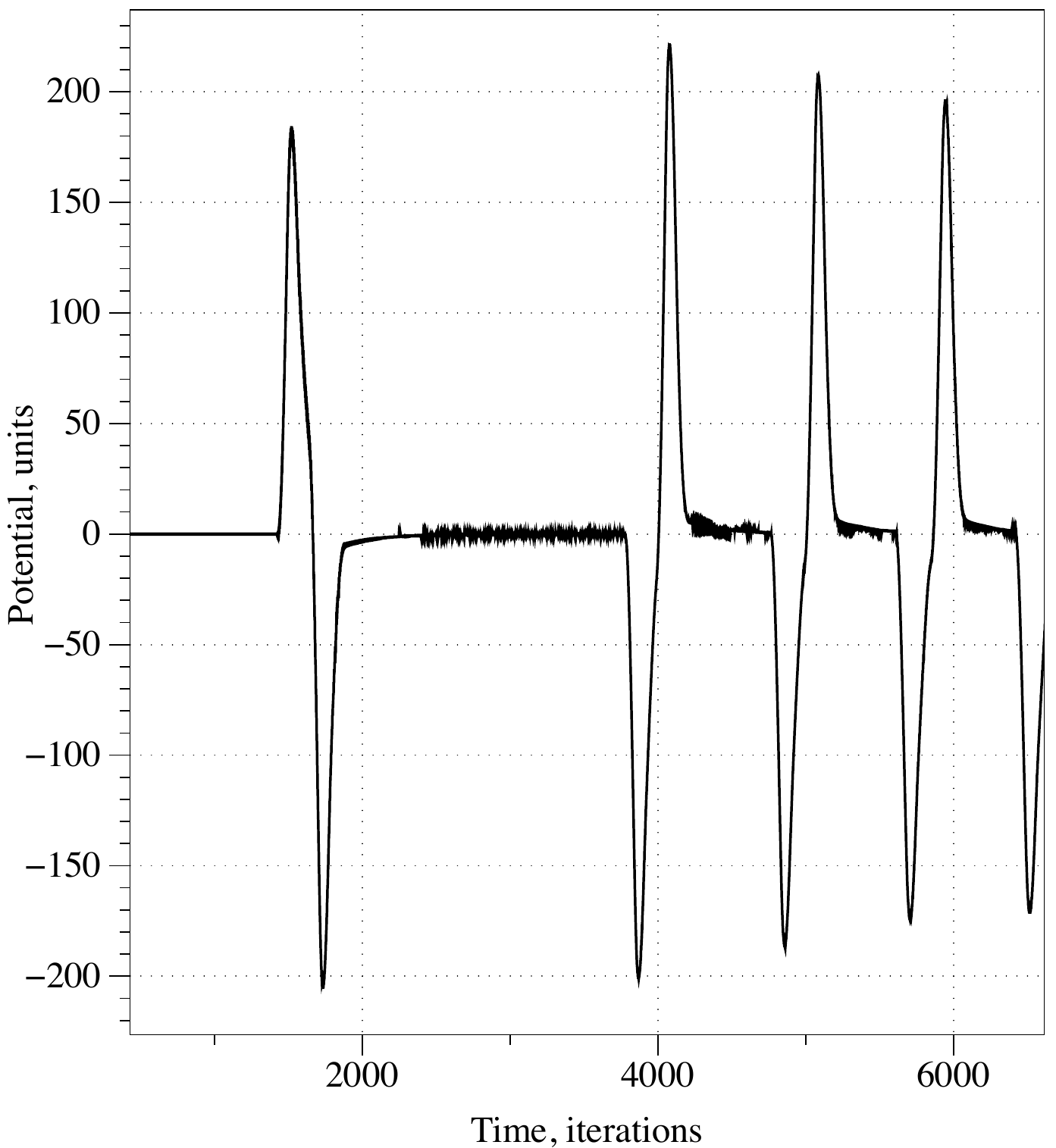}\label{fig:RareOscillationsPotential}}
    \caption{Modelling of low frequency oscillations of electrical potential in a BZ marble. (a--h)~Snapshots of excitation wave in modelled BZ marbles: electrodes are grey rectangles at the top of the black disc, oxidation wave-front is read, reduction tail is blue. (i)~Electrical potential recorded in the simulated marble. See video in \cite{adamatzkyZenodoAugust2019}.}
    \label{fig:SimulationOfRareOscillations}
\end{figure}

An excitation wave front in BZ medium has a positively charged head and negatively charged tail. Based on the shape of signals in the exemplar shown in Fig.~\ref{fig:exampleLowFrequency}, we can claim that the first spike recorded an event when the oxidation wave-front crossed electrodes in the direction from the recording to the reference electrode (thus we have sharp increase of the potential followed by the decrease). Then, an excitation wave-front crossed electrodes in the direction from the reference to the recording electrode (last three spikes in Fig.~\ref{fig:exampleLowFrequency}).

The experimental lab findings (Fig.~\ref{fig:exampleLowFrequency}) were verified in the Oregonator model, as illustrated in Fig.~\ref{fig:SimulationOfRareOscillations}. We excited the medium once on the eastern site of the marble (Fig.~\ref{fig:SimulationOfRareOscillations}a). The excitation wave first interacts with the recording electrode (Fig.~\ref{fig:SimulationOfRareOscillations}b) thus producing the ascending, depolarisation part of the potential dynamics (the first spike in Fig.~\ref{fig:SimulationOfRareOscillations}i). Then, it travels towards the reference electrode: this results in the descending, re-polarisation, part of the potential dynamics (Fig.~\ref{fig:SimulationOfRareOscillations}c). The wave then moves away from the electrode pair and the potential recorded is zero (Fig.~\ref{fig:SimulationOfRareOscillations}d).
After that we initiated a source of repeating oscillations on the `western' part of the marble (Fig.~\ref{fig:SimulationOfRareOscillations}e). The excitation wave-fronts cross the electrodes three times (Figs.~\ref{fig:SimulationOfRareOscillations}f, \ref{fig:SimulationOfRareOscillations}g and \ref{fig:SimulationOfRareOscillations}h), which results in the three latter spikes shown in the plot Fig.~\ref{fig:SimulationOfRareOscillations}i.

\begin{figure}[!tbp]
    \centering
    \subfigure[]{\includegraphics[width=0.8\textwidth]{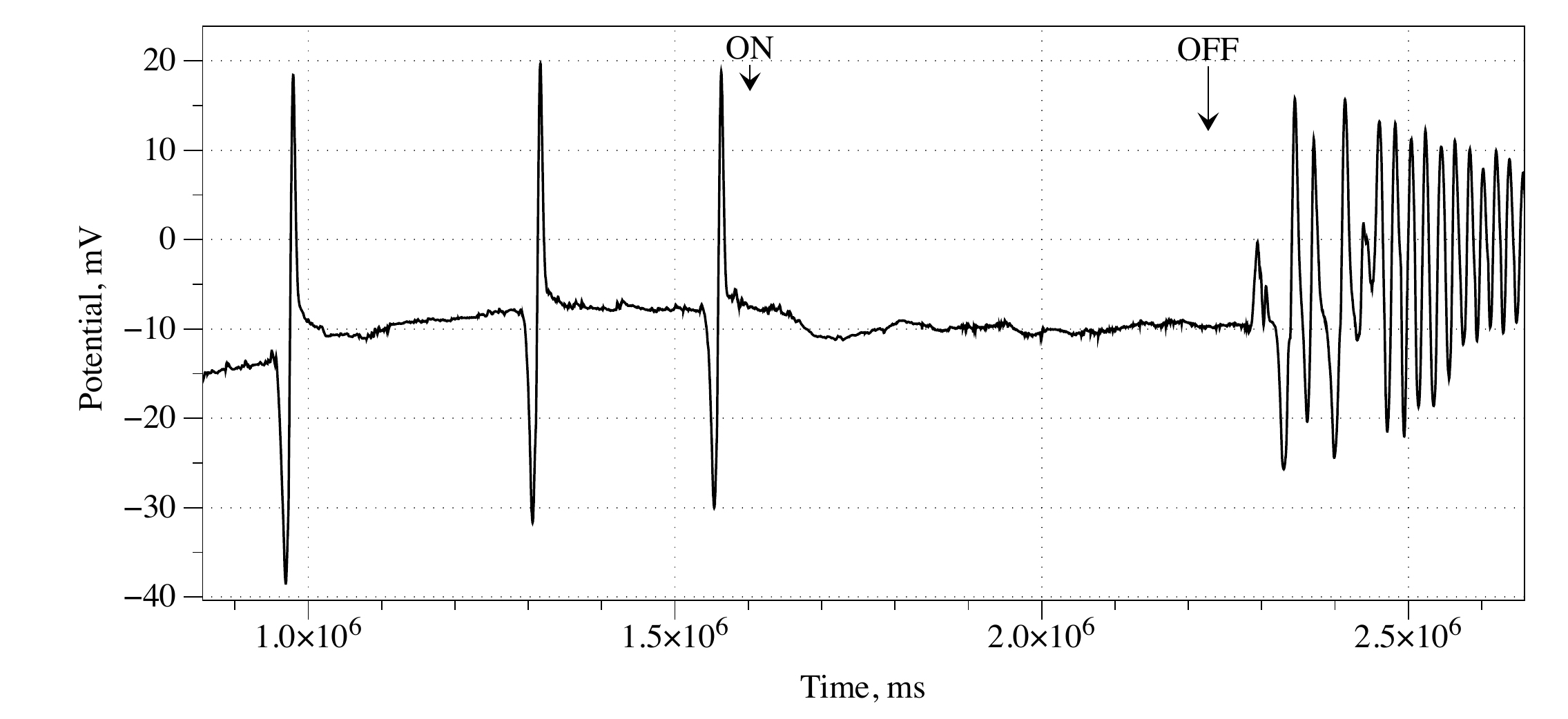}\label{fig:lowfrequencyinstantresponse}}
    \subfigure[]{\includegraphics[width=0.8\textwidth]{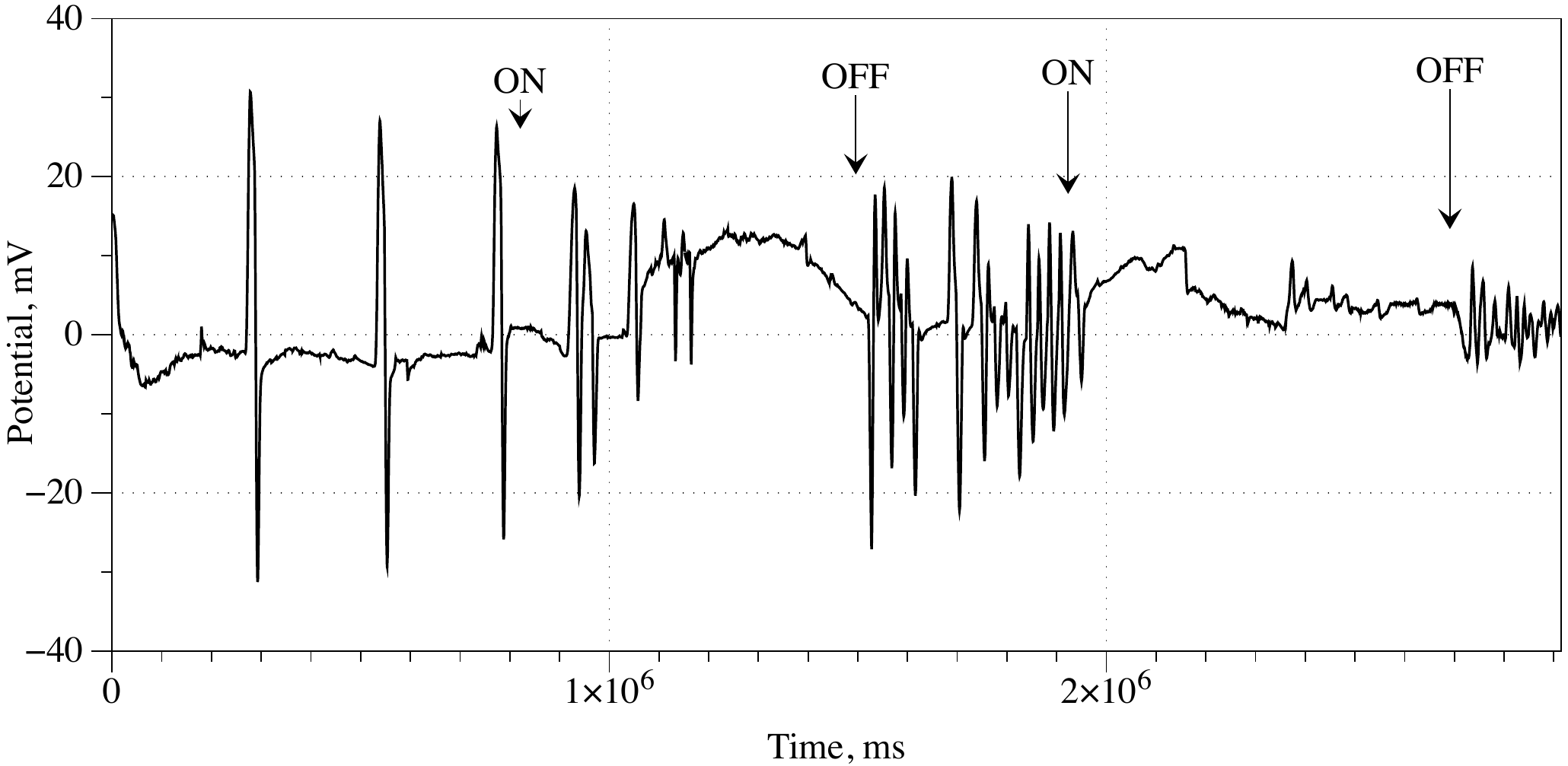}\label{fig:lowfrequencydelayedresponse}}
    \caption{Experimentally recorded low frequency response of BZ marbles to a stimulation with light. }
    \label{fig:experimentalRecordingLowFrequencyresponse}
\end{figure}

The marbles of the low frequency group reacted to the luminous stimulus as explained in the following (see exemplar potential recorded in Fig.~\ref{fig:lowfrequencyinstantresponse}). Two marbles stopped oscillating, and one marble retained its oscillations but reduced its period from 247~s to 56~s. The mean LM oscillation period after the stimulation was removed was 31~s ($\sigma=9$). Response to stimulation was instant in the two LMs that halted their oscillations; whilst the LM that continued oscillating briefly, did so in a disorganised manner for few minutes before finally halting (Fig~\ref{fig:lowfrequencydelayedresponse}). Average response to halting of stimulation for low-frequency LMs was 66~s  ($\sigma=28$). Reusability of the marble photosensor is evidenced in Fig.~\ref{fig:lowfrequencydelayedresponse} where the LM halted oscillations nearly instantly at the second round of stimulation.

\subsection{Typical frequency marbles}

\begin{figure}[htbp]
    \centering
    \subfigure[]{\includegraphics[width=0.3\textwidth]{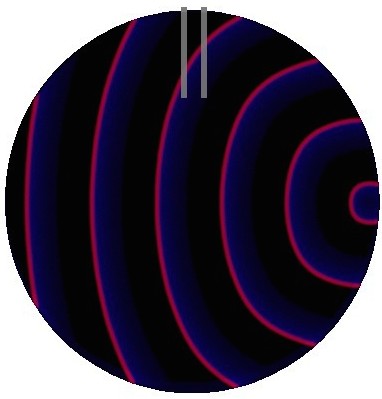}\label{fig:3ocklocksnapshot}}
    \subfigure[]{\includegraphics[width=0.5\textwidth]{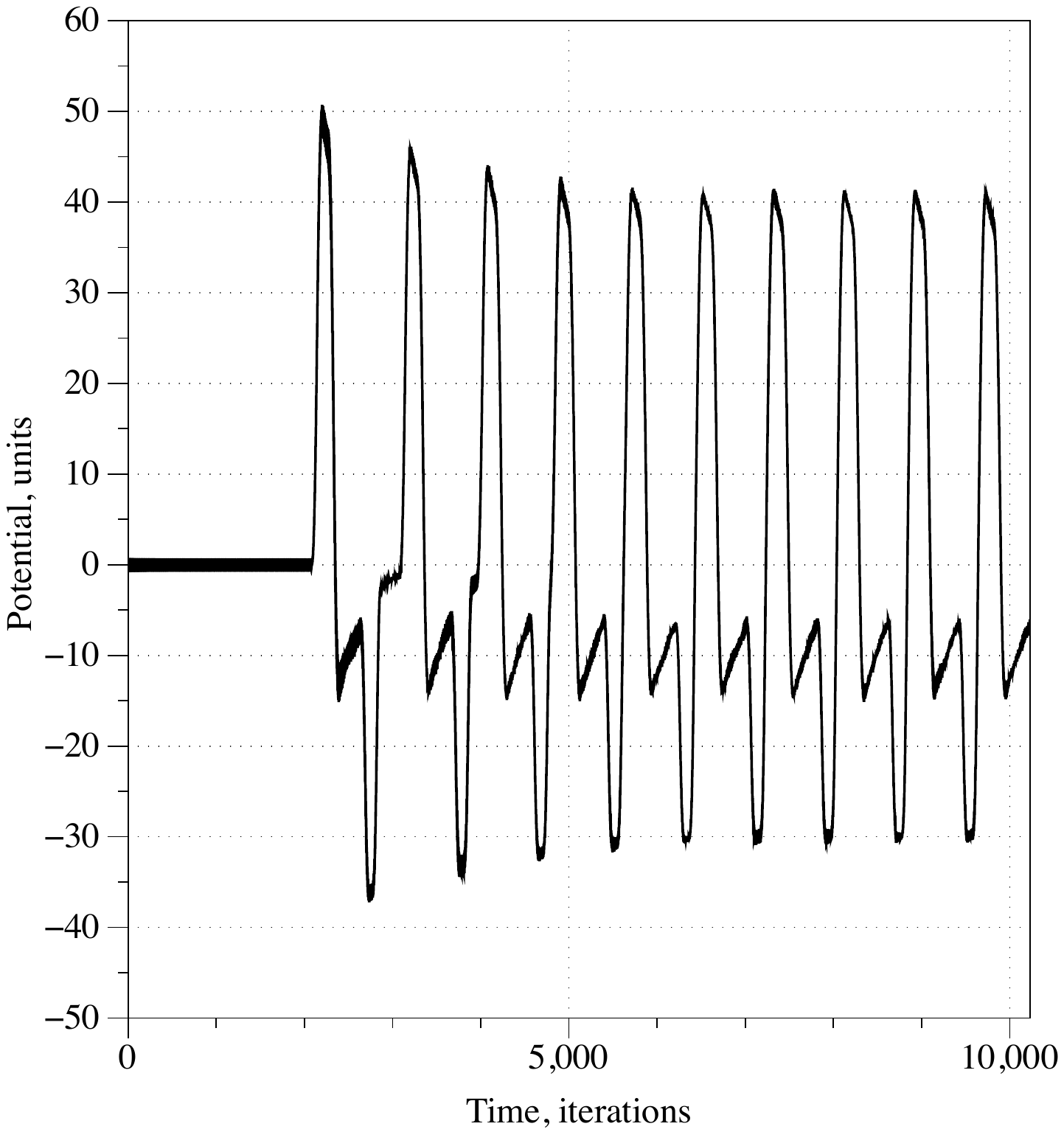}\label{fig:3ocklockpotential}}\\
    \subfigure[]{\includegraphics[width=0.3\textwidth]{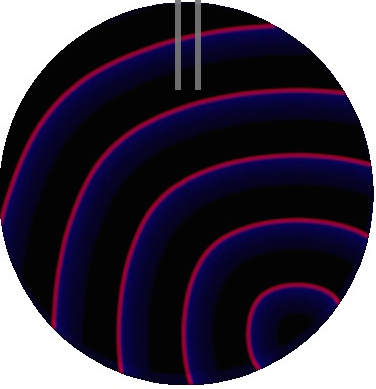}\label{fig:45ocklocksnapshot}}
    \subfigure[]{\includegraphics[width=0.5\textwidth]{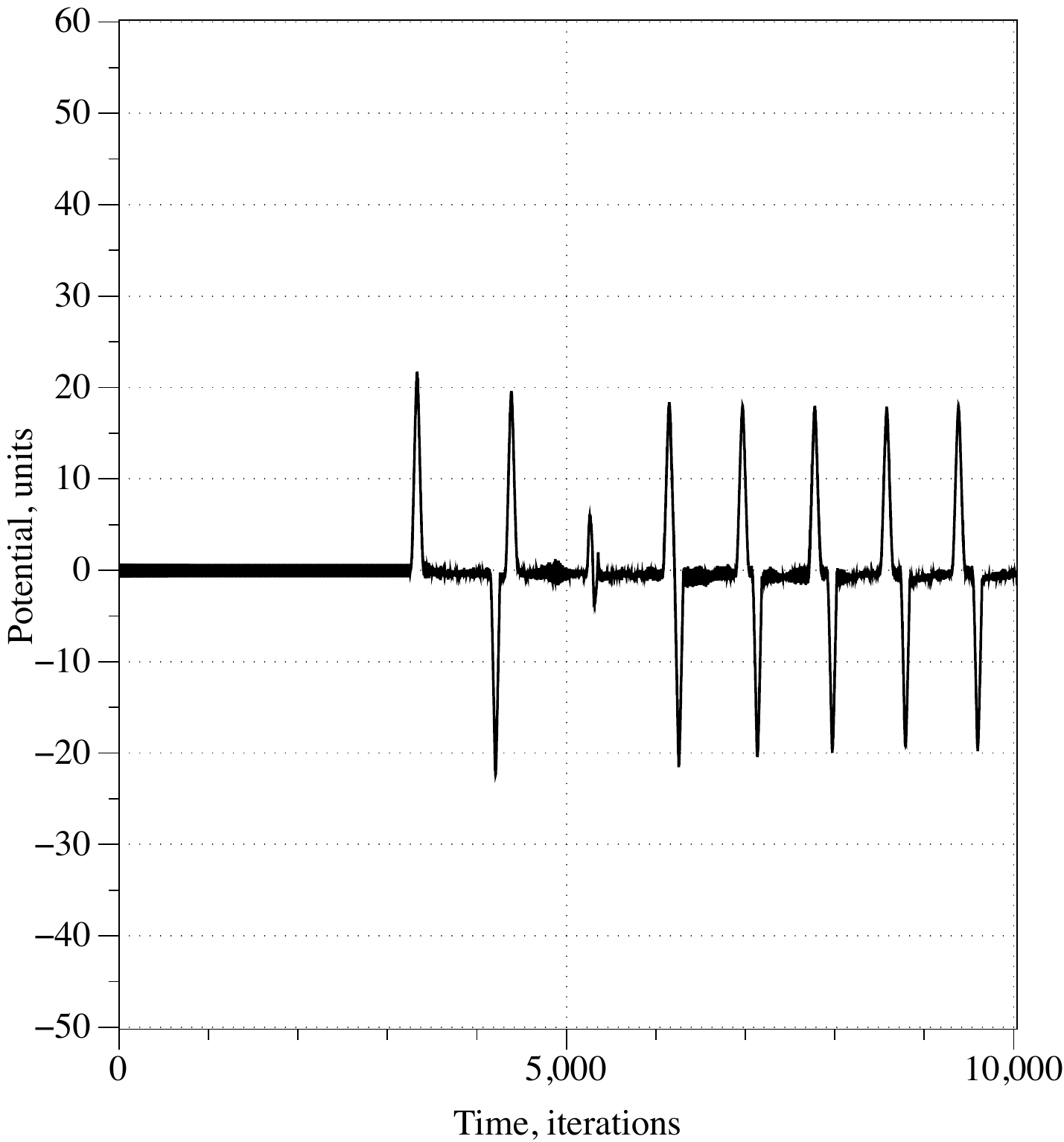}\label{fig:45ocklockpotential}}
    \caption{Pattern of oscillation of the electrical potential recorded in a BZ marble depends on the location of the source of the excitation. (a,b)~A source of excitation wave-fronts is at 3 o'clock position. (c,d)~A source of excitation wave-fronts is at 5 o'clock position. (a,c)~Snapshot of the modelled wave-fronts. (b,d)~Electrical potential recorded in the modelled marbles.
    See video in \cite{adamatzkyZenodoAugust2019}.}
    \label{fig:LocationOfSources}
\end{figure}

In the majority of cases (16 LMs, 84~\% of total), periods of oscillation (Fig.~\ref{fig:typicalperiods}) ranged from 17~s to 72~s, with an average of 36~s ($\sigma$=13) and median of 33~s, see example in Fig.~\ref{fig:exampleTypicalFrequency}. In the cases of typical oscillations, phases of the potential change are less pronounced than in low frequency oscillations. This might be due to the location of the excitation source occurring away from the electrodes. Two examples of how the location of an excitation source affects parameters of the electrical potential recorded are shown in Fig.~\ref{fig:LocationOfSources}. When the source is located at 3 o'clock position (Fig.~\ref{fig:3ocklocksnapshot}) the excitation wave-fronts approach the electrodes at an angle of approximately \SI{45}{\degree}. This is reflected in the pronounced spikes recorded (Fig.~\ref{fig:3ocklockpotential}). Source positioned at 5 o'clock (Fig.~\ref{fig:45ocklocksnapshot}) generates waves that approach the electrodes at an angle of approximately \SI{80}{\degree}. This leads to a reduced difference in electrical potential between the electrodes, reflected in the modelled signal spikes shown in Fig.~\ref{fig:45ocklockpotential}.

\begin{figure}[!tbp]
    \centering
    \includegraphics[width=0.8\textwidth]{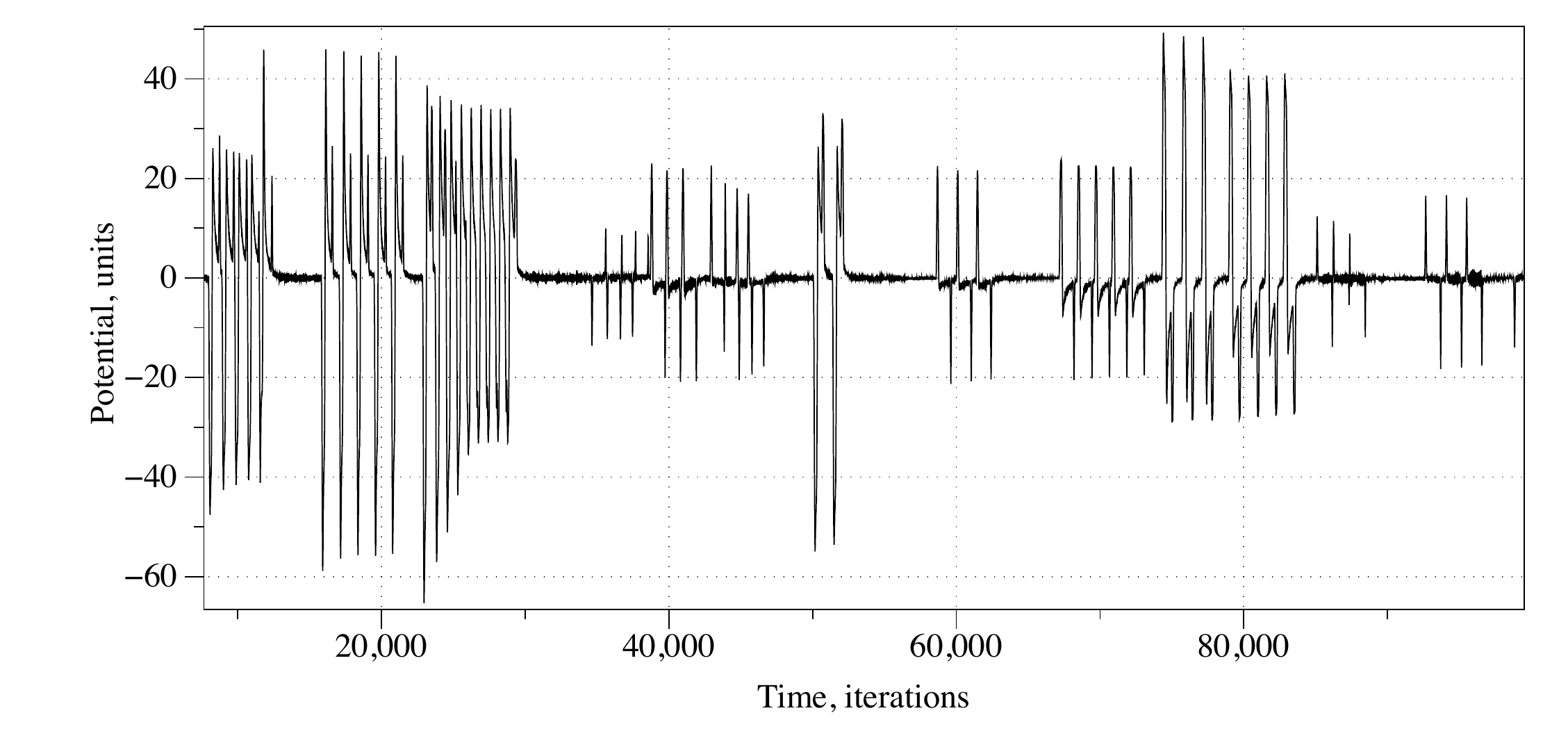}
    \caption{Modelling of bursting spiking of non-stimulated marble. There is a single source of excitation waves in the marble at each moment of time. The source has a period randomly chose from the interval $[100,700]$ and lifetime chosen from the interval $[1300,6300]$. When the current source reaches end of its lifetime it is replaced by a new source with randomly chosen parameters. See video in \cite{adamatzkyZenodoAugust2019}.}
    \label{fig:bursting}
\end{figure}

The dependence of the recorded spike waveforms, on the relative location of the excitation sources, is manifested in apparent bursts of signal spikes in the unstimulated LMs. Experimentally observed bursts are seen in Fig.~\ref{fig:averageperiodsIntact} and Fig.~\ref{fig:experimentalRecordingLowFrequencyresponse}. We can model this phenomenon by choosing the location of a source at random, allowing the source to oscillate with an arbitrary period, and allocating a short life time. When one source of excitation waves dies it is replaced by another source with randomly chosen parameters and location. The simulated signal bursts are shown in Fig.~\ref{fig:bursting}.

\begin{figure}[!tbp]
    \centering
\subfigure[]{\includegraphics[width=0.49\textwidth]{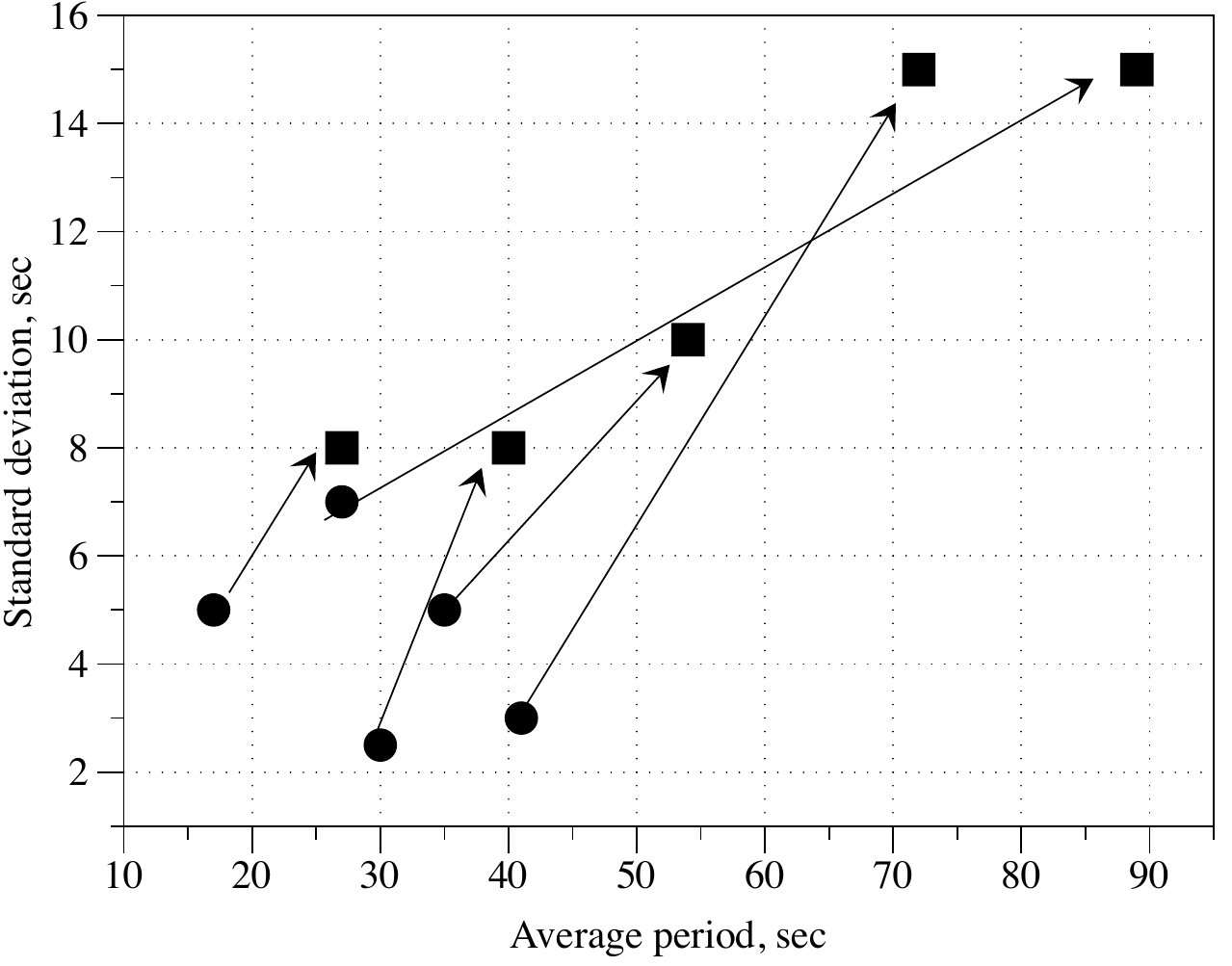}\label{fig:FrequencyMigrationDecrease}}
\subfigure[]{\includegraphics[width=0.49\textwidth]{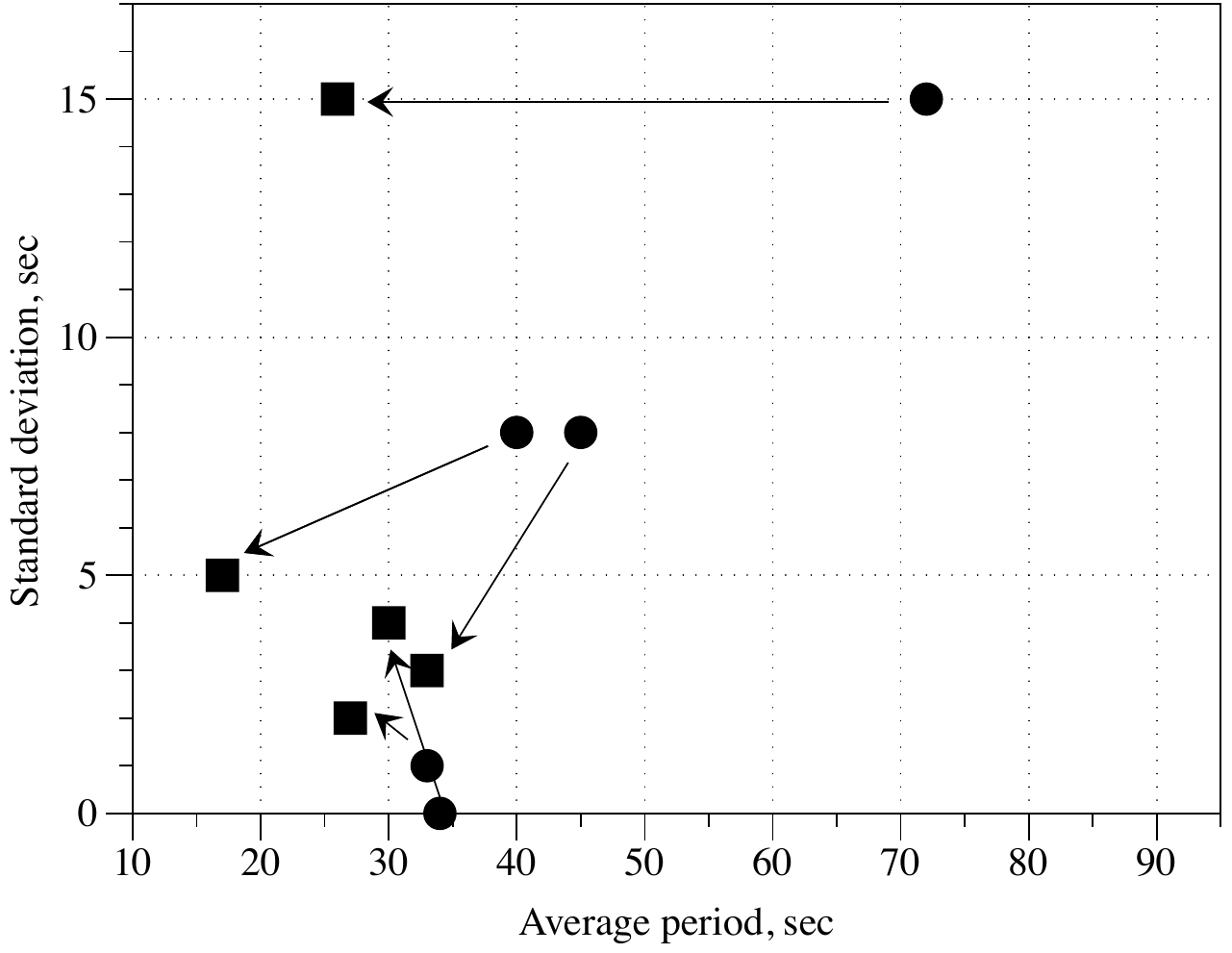}\label{fig:FrequencyMigrationIncrease}}
    \caption{Migration of marbles in the Average period of oscillations versus standard deviation. (a)~Marbles which decrease their frequency of oscillation after the stimulus switched off. (b)~Marbles which increase their frequency of oscillation after the stimulus switched off. Black discs symbolise the parameters before stimulation, black squares --- the parameters after stimulation.}
    \label{fig:FrequencyMigration}
\end{figure}

\begin{figure}[!tbp]
    \centering
\subfigure[]{\includegraphics[width=0.8\textwidth]{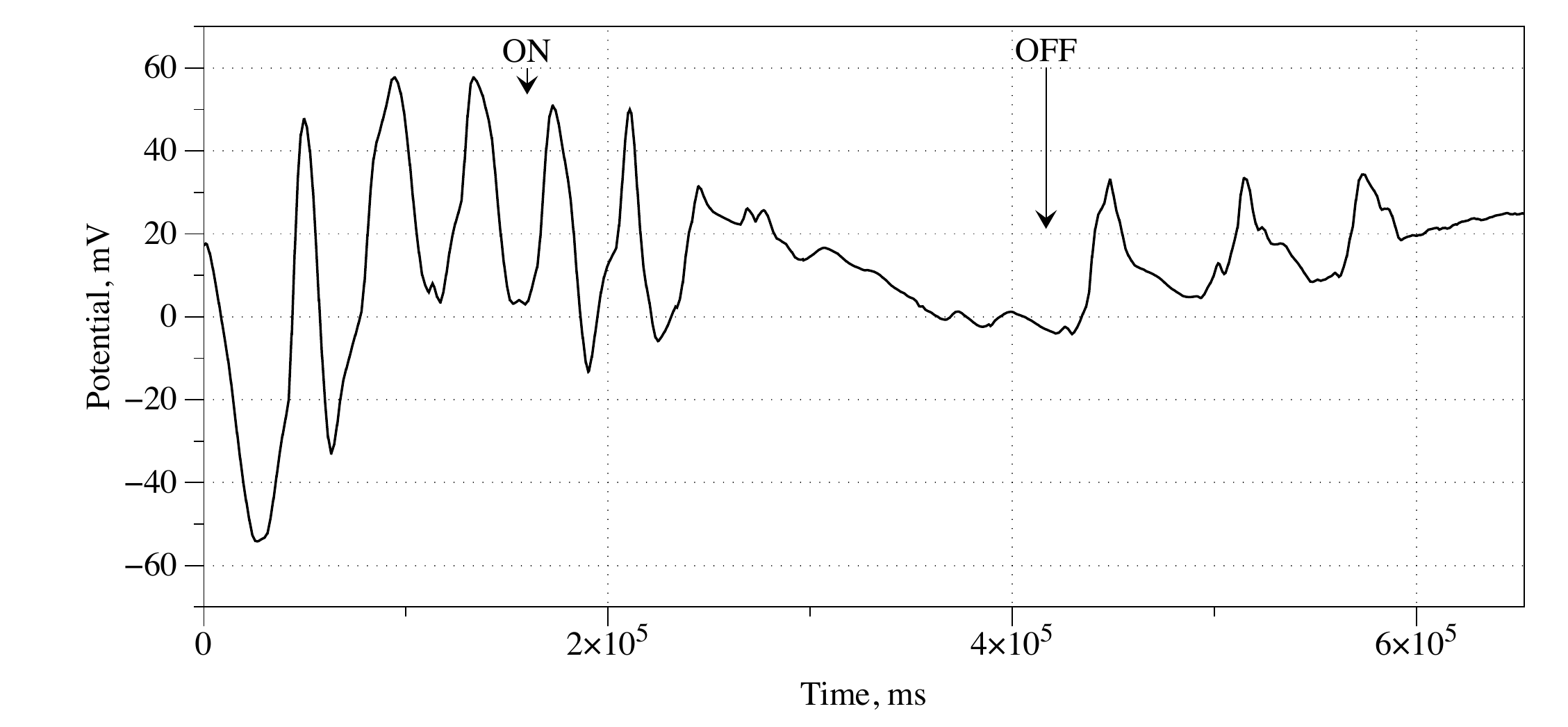}\label{fig:20June_3_DecreaseFrequency}}
\subfigure[]{\includegraphics[width=0.8\textwidth]{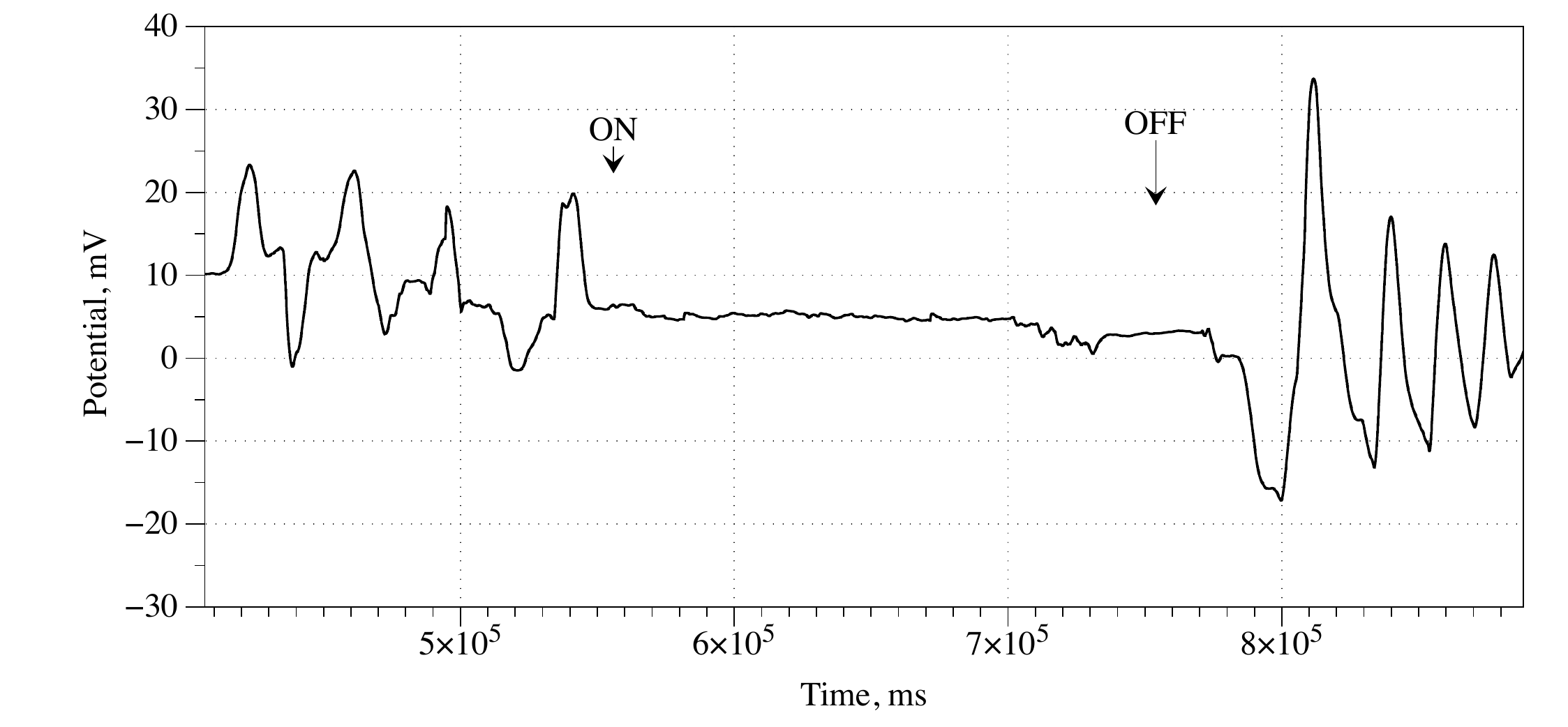}\label{fig:9July_2_IncreasedFrequency}}
    \caption{Examples of experimental marble response to stimulation in (a)~Group A and (b)~Group B.}
    \label{fig:ExampleGroupAGroupB}
\end{figure}

All marbles responded to a light stimulation with nearly instant halting of the oscillations. On switching the illumination off, the LM resumed their oscillatory activity.  The LMs can be split into two groups based on changes of the frequency after the stimulus is removed: (Group A) oscillate with decreased frequency post-stimulation (Fig.~\ref{fig:FrequencyMigrationDecrease}), six LMs, (Group B) oscillate with increased frequency (Fig.~\ref{fig:FrequencyMigrationIncrease}), eight LMs. Examples of the potential dynamics for each group are shown in Fig.~\ref{fig:ExampleGroupAGroupB}.
In Group A, the average period is changed from 30~s ($\sigma=9$) to 56~s ($\sigma=25$). In Group B, the average period is changed from 45~s ($\sigma=16$) to 27~s ($\sigma=6$). Roughly, we can say that the marbles in the Group A halve their frequencies as a result of stimulation and the marbles in the Group B double their frequencies.
The marbles respond to removal of the stimulus with nearly the same delay: Group A 41~s ($\sigma=19$) and Group B 42~s ($\sigma=30$).



\section{Summary}
\label{summary}

We demonstrated that a LM with BZ solution as the cargo and PE as the coating responds to stimulation by strong illumination by ceasing their oscillatory activity. This response is detected through changes of electrical potential difference between two electrodes inserted into a LM. The response time of the BZ marble photosensor is nearly instant, the recovery time is approximately 40~s. The lifetime of the photosensor under our experimental conditions was up to 1 hour. The sensor is reusable during its lifetime and will respond to a series of repeated stimulation. Due to their insulated coating, BZ LM photosensors can be assembled in tight clusters to act as an array of photosensors for optical inputs in a large scale chemical liquid electronics circuits. Further studies will focus on integration of the BZ photosensors into unconventional computing circuits, with optical inputs and electrical outputs.

\section{Acknowledgement}

This research was supported by the EPSRC with grant EP/P016677/1 ``Computing with Liquid Marbles''. The authors thank Dr Claire Fullarton for kindly preparing stock solutions of BZ medium and Dr Benjamin de~Lacy~Costello for originally proposing to make liquid marbles with BZ medium. A.A. thanks Jitka \v{C}ejkov\'{a} (University of Chemistry and Technology, Prague, Czech Republic) for introducing him to liquid marbles in 2016.

\end{document}